**Linton Stereo Illusion: Response on Johnston (1991)**
Paul Linton (3rd October 2024)

In (Linton, 2024) I present a new illusion (the 'Linton Stereo Illusion') that challenges our understanding of stereo vision. A vision scientist has shared their own analysis of the 'Linton Stereo Illusion' (titled: 'There is no challenge to our understanding of stereo vision: Response to Linton and Kriegeskorte (ECVP 2024 and ArXiv:2408.00770)') claiming that the 'Linton Stereo Illusion' is fully explained by Johnston (1991). I regard Johnston (1991) as one of the most important stereo vision papers in our young (< 200-year-old) field, and so this challenge requires a response. In this paper I explain why Johnston (1991) cannot explain the 'Linton Stereo Illusion'. Indeed, Johnston (1991) makes predictions that are the exact opposite of those observed in the 'Linton Stereo Illusion'. I also highlight a key concern with Johnston (1991)'s account that has so far been overlooked. Johnston (1991)'s account predicts that vergence eye movements will cause massive stereo distortions, leading to a world of unstable stereo perception. But this simply does not reflect our visual experience.

**Linton Stereo Illusion: Response on Johnston (1991)**
Paul Linton (3rd October 2024)

A vision scientist[1] has kindly asked me to clarify why I don't believe the Linton Stereo Illusion (Linton, 2024) can be explained by the analysis in (Johnston, 1991). Specifically, in (Linton, 2024) I write:

> *"Could this effect be explained by (Johnston, 1991)? No. Johnston assumes disparities are scaled with a distance estimate of ≈80cm, so Johnston's account would predict that in the 'Constant Physical Separation' condition we see the far separation, not the near separation, as accentuated in depth."*

To give some background, (Johnston, 1991) concludes that binocular disparities are 'scaled' by the visual system using an inaccurate estimate ($y$) of the viewing distance ($x$), where:

$$y \approx 0.265x + 58.8$$

Which is more usefully reformulated as:

$$y \approx 80 + 0.265*(x - 80)$$

Weak Triangulation ⟵ ⟶ Strong Triangulation

We can think of this formula as having two components:

1. <u>Abathic Distance (Weak Triangulation)</u>: A default internal estimate of viewing distance, known as the 'abathic distance', that is roughly 80cm (where $y = x$).

2. <u>Vergence (Strong Triangulation)</u>: A refinement of the 'abathic distance' based on a faulty (gain of 0.265) estimate of $x$, thought to come from 'vergence'.

I call these 'Weak Triangulation' and 'Strong Triangulation' because both are classic triangulation accounts that date back to (Kepler, 1604) and (Descartes, 1637), the only question is whether they use a default internal estimate of the viewing distance or try to actively estimate it using vergence.

Why does this matter? Well, because in (Linton, 2023) I challenge *all* triangulation accounts of stereo vision, arguing that perceived stereo depth is simply a function of the disparities on the retina. However, my experimental work has so far only focused on challenging vergence as a distance (Linton, 2020) and size (Linton, 2021) cue, leaving the core of 'Weak Triangulation' untouched. The Linton Stereo Illusion is therefore meant to be a way of challenging both Weak and Strong Triangulation.

**1. Modelling Vergence Fixed at 40cm in 'Constant Physical Separation' Condition**

To see why, let's start with (Johnston, 1991)'s own experimental paradigm: testing stereo shape perception at 3 fixed vergence distances: close (53.5cm), medium (107cm), and far (214cm), by having participants manipulate the depth of a stereo cylinder so it looked 'regular' (depth = ½ height).

---

[1] Personal communication, 'There is no challenge to our understanding of stereo vision: Response to Linton and Kriegeskorte (ECVP 2024 and ArXiv, https://arxiv.org/abs/2408.00770)', 9th September 2024.

What (Johnston, 1991) found was close to veridical 3D shape perception at medium distances (estimated to be 80cm), over-constancy at close distances (< 80cm) and under-constancy at far distances (> 80cm). So, the disparity defined cylinders that looked regular at 53.5cm and 214cm were:

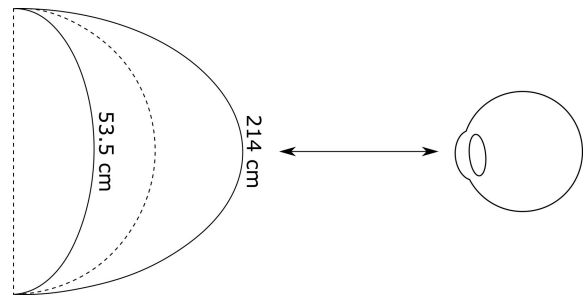

Figure 1: The 3D shape of cylinders perceived to be 'regular' by observers in (Johnston, 1991) at near (53.5cm) and far (214cm) viewing distances.

But there is another explanation for these results (or something close to them), namely that these distortions simply reflect how disparities fall off with the viewing distance squared. So something like these results would also be predicted by (Linton, 2023)'s 'Minimal Model of 3D Vision'.

So how can we go about differentiate between these two accounts? One of the key insights of (Linton, 2023) is that whilst this is relatively difficult for stimuli with two depth-planes (a front and a back, like (Johnston, 1991)'s cylinders), the Triangulation and Disparity (or Minimal) models make very different predictions about three depth-plane stimuli (see (Linton, 2023), fig.9 and associated text). It is this insight which is exploited in the Linton Stereo Illusion as a way of differentiating between the Triangulation and Disparity (or Minimal) models. Consider three points equally spaced in depth:

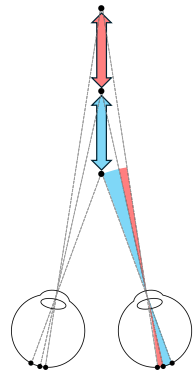

Figure 2. Equal physical distances in depth lead to unequal disparities on the retina.

The Linton Stereo Illusion relies on two key insights: First, it doesn't matter what the (equal) separation between the dots is, or how far the dots are away from the observer, geometry dictates that the closer separation (in blue) will project larger on the retina than the further separation (in red). So, in order to get back to the equal physical separation from the retinal projections, the disparities of the further separation (in red) have to be <u>scaled more</u> than the disparities of the nearer separation (in blue).

The second insight is how scaling distance affects this process. Consider the Linton Stereo Illusion <u>if the vergence distance is kept constant at 40cm (the surface of the display) throughout</u>. What happens if the disparities in the Linton Stereo Illusion are scaled by an internal estimate of distance greater than 40cm, as (Johnston, 1991) suggests? (Johnston, 1991)'s formula specifies a scaling distance of ≈70cm, and Fig.2 shows how increasing the scaling distance from 40cm to 70cm changes our percept:

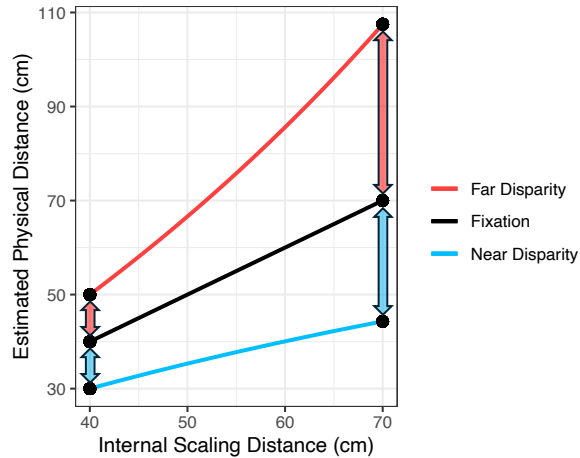

Figure 3. Disparities of the Linton Stereo Illusion replotted for different scaling distances.

Two things should immediately become apparent when we compare $x = 40$cm to $x = 70$cm in Figure 3, both of which count strongly against the (Johnston, 1991) interpretation:

1. <u>Implausible Absolute Distances</u>: First, whilst the three points originally correspond to:

Near Disparity: **30cm** (red), Fixation: **40cm** (black), and Far Disparity: **50cm** (blue)

Now, when rescaled using (Johnston, 1991)'s estimate of 70cm, they are transformed to:

Near Disparity: **44cm** (red), Fixation: **70cm** (black), and Far Disparity: **107cm** (blue)

This means that that the near disparity should be perceived as having 26cm (70cm – 44cm) of depth. This is a very literal prediction of the (Johnston, 1991). It's common to think of (Johnston, 1991) as just a paper about perceived 3D shape from stereo. But the way (Johnston, 1991) measures this perceived geometry is by matching the z-axis depth to a y-axis length. Presuming we have a roughly veridical idea of x-axis and y-axis lengths on a screen 40cm away in full cue conditions, which I don't think anyone is questioning, then it follows that the perceived depth off the screen must be 26cm.

The problem, of course, is that the screen itself is 40cm away. This means that the near disparity must be perceived as being 14cm from our face (40cm – 26cm). This is implausibly close. But in any case, you can test it for yourself using the anaglyph version of the Linton Stereo Illusion by holding a finger out 14cm from your face. The Linton Stereo Illusion seems nowhere close to the predicted 14cm.

2. <u>Implausible Geometry</u>: However, the key insight that the Linton Stereo Illusion seeks to exploit is the second implausibility of the (Johnston, 1991) interpretation of the Linton Stereo Illusion.

The key insight is realizing what (Johnston, 1991)'s near-distance 'over-constancy' means for the two separation staggered in depth (as in Figure 2). As we see in Figure 3, when rescaled using 70cm as the scaling distance, rather than the two separations being perceived as equal (as they are physically), now the further separation is perceived as larger (37cm) than the near separation (26cm). The ratio near:far is 1:1.4 (far > near). But the key observation of the Linton Stereo Illusion in the Constant Physical Separation condition is that the near disparity is perceived as being larger than the far disparity (near > far), the exact opposite of what is predicted by (Johnston, 1991)'s account. I therefore wrote:

> *"Could this effect be explained by (Johnston, 1991)? No. Johnston assumes disparities are scaled with a distance estimate of ≈80cm, so Johnston's account would predict that in the 'Constant Physical Separation' condition we see the far separation, not the near separation, as accentuated in depth."* (Linton, 2024)

This is an important point, and the intuition behind it is crucial to understanding any account (such as (Johnston, 1991)'s) that rests on an 'abathic distance'. Start with stereo under-constancy at far distances, something we experience in everyday viewing conditions: stereo depth is increasingly weak with distance, indicating that further separations in depth are under-scaled relative to near separations in depth. Now the logic of 'abathic distances' is that the abathic distance is a flipping point, closer than which, the exact opposite distortions should be experienced. So whilst Triangulation ('abathic distance') models and Disparity (or Minimal) models make very similar predictions about two separations in depth viewed at far distances, Triangulation ('abathic distance') accounts predict these distortions should become 'flipped' at near distances, a claim Disparity (or Minimal) models reject.

You may respond by saying that vergence was not fixed in my ECVP and VSS demonstrations. That is true. But you can retest the illusion for yourself using the anaglyph version. Set the screen at 40cm, fix your vergence on the screen, and watch the illusion. The separation between the circles appears to expand as it gets closer (as predicted by my account) not compress (as predicted by (Johnston, 1991)).

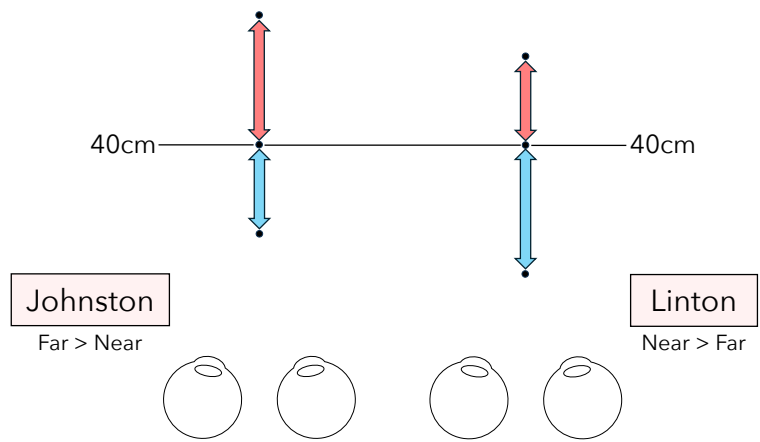

Figure 4. Competing predictions of (Johnston, 1991) and (Linton, 2023) for the Constant Physical Separation condition of the Linton Stereo Illusion, with vergence fixed at 40cm.

For me, this should be the end of the discussion. With fixed vergence, (Johnston, 1991)'s account makes predictions which are immediately shown to be both implausible in the case of Absolute Distance and the exact opposite of what is actually perceived in the case of Geometry.

**2. Modelling Vergence Tracking of the Near Circle in 'Constant Retinal Disparity' Condition**

The vision scientist raises the following scenario[2]. Consider the Fixed Disparity condition of the Linton Stereo Illusion. Why is that perceived as being rigid? On my account, it's because the disparities between the two circles don't change. But they argue that (Johnston, 1991), too, can account for this. To see why, consider the Weak and Strong Triangulation aspects of (Johnston, 1991):

$$y \approx 80 + 0.265*(x - 80)$$

Weak Triangulation ⟶ ⟵ Strong Triangulation

Imagine we only had the Weak Triangulation component. As vergence tracks the near circle[3], and the disparity of the far circle remains fixed, the separation in depth is always perceived to be the same because it's scaled using the same 80cm estimate of viewing distance no matter the vergence distance.

Furthermore, they makes the point that given the small vergence gain (0.265), the change in perceived depth implied by the addition of Strong Triangulation over 10cm should be relatively small.

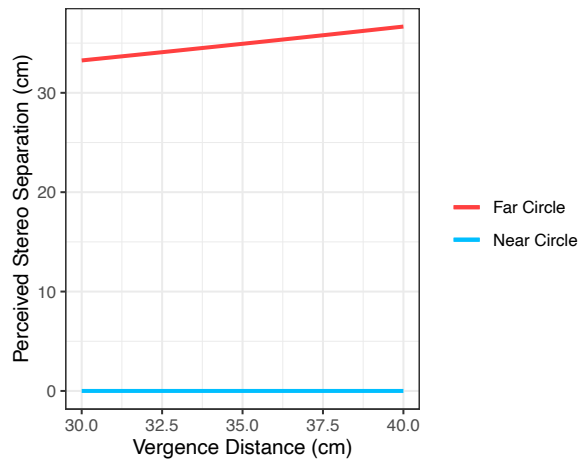

Figure 5. (Johnston, 1991)'s prediction of how the perceived stereo depth between the two circles should change as vergence tracks the near circle from 30cm to 40cm.

However, if we consider Figure 5, the same two concerns we raised before again become immediately apparent, and both of them count strongly against the (Johnston, 1991) interpretation:

---

[2] Personal communication, 'There is no challenge to our understanding of stereo vision: Response to Linton and Kriegeskorte (ECVP 2024 and ArXiv, https://arxiv.org/abs/2408.00770)', 9th September 2024.
[3] Roughly the same logic applies to tracking the far circle rather than the near one. But since near circle largely occludes far one, tracking near circle is the case I will consider here as the most likely of the two scenarios.

1. <u>Implausible Absolute Distance</u>: The predicted perceived distance of the separation ranges from 33.25cm to 36.67cm. First, this is implausibly large compared to our actual experience of the Linton Stereo Illusion. Again, you can try for yourself using the anaglyph version. Second, this is impossibly large for the 'near' position (vergence = 30cm), since the screen is perceived to be 40cm away in full cue conditions, meaning that the front circle can only be 6.75cm (40cm – 33.25cm) from the observer.

2. <u>Debate Over Geometry:</u> They are right that the change in the perceived size of the separation is relatively small (a reduction of 10%) as vergence moves from 40cm (separation = 36.7cm) to 30cm (separation = 33.3cm). Their point being that the two lines are not a million miles away from parallel. But first, a change in 10% will likely be detectable. And second, we risk losing sight of the fact that this change corresponds to a change of 3.4cm, which isn't at all evident in our perception of the Linton Stereo Illusion (and gets washed out in Fig.5 by the large absolute distances involved).

**3. Modelling Vergence Alternating Between the Near and Far Circles (at t = 0)**

However, (Johnston, 1991)'s rescaling of disparities with a changing vergence is not as innocuous as it first appears. Whilst changing vergence does not have a massive effect on perceived depth in Fig.5, this is because the disparities always remain further than fixation. But now consider the case where we alternate fixation from near (40cm) to far (50cm). This is the scenario involved in alternating fixation from near to far in the Linton Stereo Illusion at t = 0, where both the Constant Physical Separation and Constant Retinal Disparities conditions start with a near circle at 40cm and a far circle at 50cm.

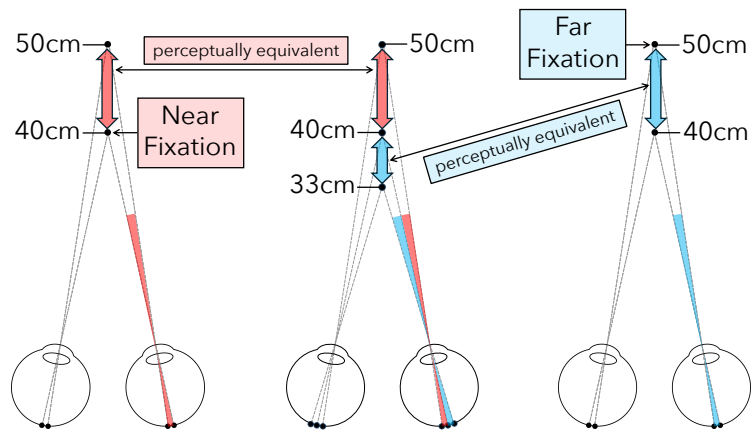

Figure 6. The very same disparities are interpreted as having different depths under (Johnston, 1991)'s account if the true vergence distance is (largely) ignored.

The important point to realize is that if (Johnston, 1991) is correct, and the scaling distance is effectively the same (≈ 80cm) no matter what the real vergence distance is, then the visual system will interpret near disparities from 50cm fixation (Fig.6 right, blue) the same as it would interpret near disparities from 40cm (Fig.6 centre, blue), even though the visual system will interpret those very same disparities as far disparities from 40cm (Fig.6 centre, red) when fixation shifts to 40cm (Fig.6 left, red).

Put simply, (Johnston, 1991)'s account predicts the perceived depth of the very same disparities will distort (difference between red and blue arrows in Fig.6 centre) depending on whether we fixate near (Fig.6 left, red) or fixate far (Fig.6 right, blue). Now this account requires a little bit of refinement, because (Johnston, 1991) argues that vergence does in fact weakly affect the scaling distance (with a gain of 0.265). But the reason I frame the point in this way is to emphasise that what seemed to be a strength of (Johnston, 1991)'s account in trying to explain the Linton Stereo Illusion in Section 2, leads to some pretty significant (and pretty damning) unintended consequences down the road.

Indeed, even with the vergence gain of 0.265 incorporated, (Johnston, 1991)'s account predicts a separation of 10cm between 40cm and 50cm will be perceived as 36.7cm in depth when we fixate near (at 40cm), but only 19cm when we fixate far (at 50cm). Put simply, the apparent stereo depth of the separation should expand by ≈50% as we shift our fixation from far (50cm) to near (40cm).

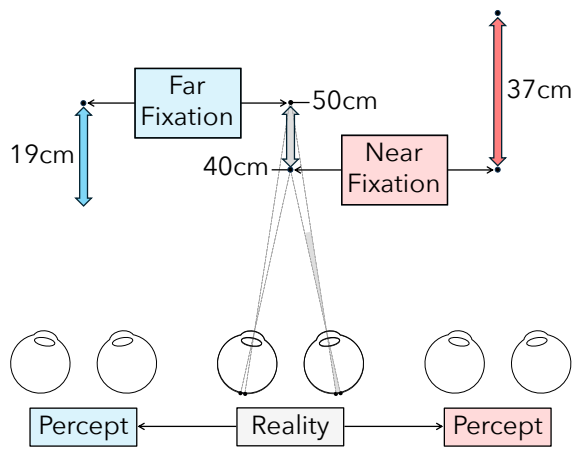

Figure 7. Illustration of how the perception of the very same physical distance distorts with near and far fixation under (Johnston, 1991)'s account. Note that the absolute size of the percepts are not to scale. As I've emphasised throughout, (Johnston, 1991)'s account predicts implausibly large absolute distances. The key point being conveyed here is that (Johnston, 1991)'s account predicts that the stereo depth percept should expand by ≈50% as we shift our fixation from far (50cm) to near (40cm).

**Do Eye Movements Distort Stereo Depth?**

Ironically, the one thing that traditional Triangulation accounts with full depth constancy (scaling with $y = x$) and my Minimal account (direct access to disparity) have in common is that the perceived depth of a 3D scene is invariant to changes in eye movement[4], either because there is no re-scaling of disparities with each eye movement (my account), or because the scaling distance used to re-scale disparities with each eye movement is veridical ($y = x$). And this coheres with our everyday experience of the world. Fixating near, and then fixating far, does not massively distort the stereo depth in the scene. Put simply, there is not a problem of 'unstable stereo depth' caused by changing vergence.

The problem comes when we try to maintain an intermediate (and, I would argue, unstable) halfway house between these two positions, such as (Johnston, 1991)'s. As we saw in Section 3, when the disparities in the scene are improperly re-scaled with each eye movement ($y \neq x$), these lead to massive distortions in our perception of stereo depth. On such an account, as we move our eyes around the scene, the world should constantly distort in stereo depth, leading to the percept of an unstable (constantly warping) 3D world. But this simply doesn't reflect our actual everyday visual experience.

---

[4] There will be some very subtle changes to the disparities on the retina with vergence eye movements ('gaze contingent disparities': (Turski, 2016), (Linton, 2019)) due to the slight offset of the nodal point and the centre of rotation of the eye. But the key point is that retinal disparities are not being re-scaled as having different perceived depths.

It also explains why there's a debate between me and the vision scientist over (Johnston, 1991)'s ability to explain the Linton Stereo Illusion in the first place. The key difference is that I analyse the Linton Stereo Illusion with fixed vergence (Section 1), whilst they analyses it with changing vergence (Section 2). That two leading stereo vision scientists can come to such disparate conclusions about the same stimulus, and both claim victory, simply on the basis of different assumptions about vergence dynamics is an indication that something is seriously wrong with the (Johnston, 1991) account.

This shortcoming of the (Johnston, 1991) account rests in particular on missing the significance of the fact that the very same disparities will be re-scaled as 'near' or 'far' disparities depending on where on the stimulus (front or back) fixation is, leading to very different magnitudes under a triangulation account that doesn't scale using a perfect internal estimate of the viewing distance. But why has this all-important near-far dynamic been (so far as I am aware) overlooked in the literature? After all, (Johnston, 1991) has been the leading account of disparity scaling for the last 33 years.

Part of the reason might be our impoverished stereo stimuli. Traditionally, the literature has focused on testing simple (two-plane) stimuli with just a 'front' and a 'back', like (Johnston, 1991)'s cylinder. By contrast, in full-cue conditions, the Linton Stereo Illusion tests (at least) three planes: near (30cm), middle (40cm), and far (50cm), with the whole point of the Linton Stereo Illusion being to emphasise this important distinction between scaling the very same disparities as 'near' or 'far'. The three-plane configuration of the Linton Stereo Illusion was more explicitly articulated in the original version of the illusion (submitted to VSS Demo Night)[5], which had static markers at 30cm, 40cm, and 50cm. The static markers were removed when it became apparent the illusion worked just as well in full-cue conditions, and so the frame of the display functioned as the static marker for 40cm.

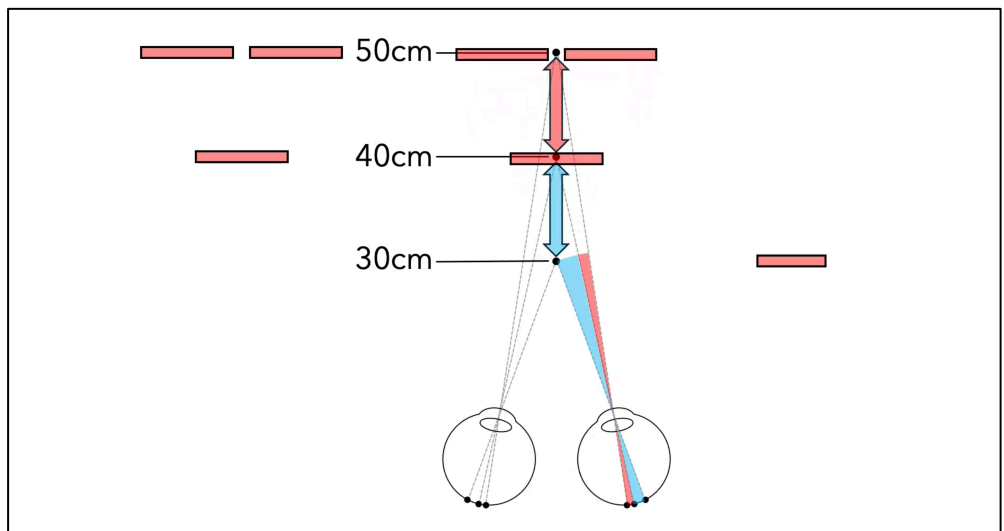

Video 1. Original configuration of the Linton Stereo Illusion.
[Insert Video 1 here: https://youtu.be/z9V4waTmae8]

---

[5] Version submitted to VSS Demo Night (17th March 2024): https://www.youtube.com/watch?v=El7YKaURsuQ

**Conclusion**

The vision scientist and I agree that (Johnston, 1991) is one of the most important stereo vision papers of the last 50 years. Indeed, I would rank it as one of the most important stereo vision papers ever in our young (< 200-year-old) field. It poses a dilemma to the field (distortion of stereo shape) that I don't believe has been adequately answered for 33 years. Part of the reason (as I document in this paper) is that we have failed to take seriously the predictions of (Johnston, 1991)'s own account. But this failure reflects a more fundamental failing of the field. For the last 150 years – from (Helmholtz, 1866) to the present[6] – we have constructed increasingly Baroque edifices out of prior theories in an attempt to account for stereo distortions, rather than embracing newer and simpler theories that provide an immediate (and almost irresistible) explanation (Linton, 2023)(Kuhn, 1969).

**References**

Descartes, R. (1637). Dioptrique (Optics). In J. Cottingham, R. Stoothoff, & D. Murdoch (Eds.), The

Philosophical Writings of Descartes: Volume 1 (1985). Cambridge University Press.

Helmholtz, H. (1866). Handbuch der Physiologischen Optik, Vol.III (translated by J. P. C. Southall

1925 Opt. Soc. Am. Section 26, reprinted New York: Dover, 1962).

Johnston, E. B. (1991). Systematic distortions of shape from stereopsis. Vision Research, 31(7),

1351–1360. https://doi.org/10.1016/0042-6989(91)90056-B

Kepler, J. (1604). Paralipomena to Witelo (W. H. Donahue, Trans.). In Optics: Paralipomena to

Witelo and Optical Part of Astronomy. Green Lion Press, 2000.

Kuhn, T. S. (1969). The Structure of Scientific Revolutions. University of Chicago Press.

Linton, P. (2019). Would Gaze-Contingent Rendering Improve Depth Perception in Virtual and

Augmented Reality? https://arxiv.org/abs/1905.10366v1

Linton, P. (2020). Does vision extract absolute distance from vergence? Attention, Perception, &

Psychophysics, 82(6), 3176–3195. https://doi.org/10.3758/s13414-020-02006-1

Linton, P. (2021). Does Vergence Affect Perceived Size? Vision, 5(3), Article 3.

https://doi.org/10.3390/vision5030033

---

[6] As (Johnston, 1991) documents, (Johnston, 1991) is the continuation of a tradition that starts with (Helmholtz, 1866) of attributing stereo depth distortions to disparity scaling with the wrong internal estimate of the viewing distance.

Linton, P. (2023). Minimal theory of 3D vision: New approach to visual scale and visual shape. Philosophical Transactions of the Royal Society B: Biological Sciences, 378(1869), 20210455. https://doi.org/10.1098/rstb.2021.0455

Linton, P. (2024). Linton Stereo Illusion (arXiv:2408.00770). arXiv. https://doi.org/10.48550/arXiv.2408.00770

Turski, J. (2016). On binocular vision: The geometric horopter and Cyclopean eye. Vision Research, 119, 73–81. https://doi.org/10.1016/j.visres.2015.11.001